\documentstyle[epsf,aps,pre,multicol]{revtex}

\begin{document}

\title{Velocity Distributions and Correlations
in Homogeneously Heated Granular Media
}
\author{Sung Joon Moon~\cite{email}, M. D. Shattuck~\cite{newaddress},
        and J. B. Swift}
\address{Center for Nonlinear Dynamics and Department of Physics,
         The University of Texas at Austin, Austin, TX 78712}
\date{\today}
\maketitle

\begin{abstract}
We compare the steady state velocity distributions from
our three-dimensional inelastic hard sphere molecular dynamics simulation
for homogeneously heated granular media, with the predictions of
a mean field-type Enskog-Boltzmann equation for inelastic hard
spheres [van Noije \& Ernst, Gran. Matt. {\bf 1}, 57 (1998)].
Although we find qualitative agreement for all values of density and
inelasticity, the quantitative disagreement approaches $\sim 40\%$
at high inelasticity or density.
By contrast the predictions of the pseudo-Maxwell molecule model
[Carrillo, Cercignani \& Gamba, Phys. Rev. E, {\bf 62}, 7700 (2000)]
are both qualitatively and quantitatively different from those of
our simulation.
We also measure short-range and long-range velocity correlations
exhibiting non-zero correlations at contact before the collision,
and being consistent with a slow algebraic decay over a decade
in the unit of the diameter of the particle, proportional to
$r^{-(1+\alpha)}$, where $0.2 < \alpha < 0.3$.
The existence of these correlations imply the failure of
the molecular chaos assumption and the mean field approximation,
which is responsible for the quantitative disagreement of 
the inelastic hard sphere kinetic theory.

\end{abstract}
\nobreak
\begin{multicols}{2}

\section{introduction}
Granular materials are collections of noncohesive macroscopic
dissipative particles and are encountered in nature
and in the industry~\cite{jaeger}.
These materials exhibit a wide variety of phenomena depending on
the external forcing.
The rapid granular flow regime, where the collisions are modeled
as instantaneous binary inelastic collisions,
is reminiscent of a gas of hard spheres.
Thus, a common theoretical approach for this regime as a first order
approximation is to model the system by means of the kinetic and
the continuum equations for smooth inelastic hard spheres with
a velocity-independent coefficient of restitution~\cite{continuum}.
In the kinetic theory approach, the mean field-type Boltzmann or
Enskog-Boltzmann equation for inelastic hard spheres is used,
and most techniques are directly transposed from the kinetic theory
of a gas of elastic hard spheres~\cite{chapman}.
It is known that this formulation is a reasonable description for nearly
elastic ($1-e^2 \ll 1$, where $e$ is the normal coefficient of restitution)
and dilute cases.
However, these theoretical models include approximations such as
the truncation of series expansion or hierarchy,
and the introduction of equation closure.
The dissipative nature of the collision modifies the physics
in a nontrivial way, and the accuracy and the limitation of the mean field-type
kinetic description with these approximations is not yet known.
The extension of the theory for the more inelastic and dense case,
including surface friction, is one of major goals of
the current inelastic kinetic theory~\cite{goldhirsh}.

There is an attractor in the phase space of the granular media,
because inelastic collisions dissipate kinetic energy;
in the absence of an external energy source, a granular medium
loses its kinetic energy through collisions and becomes a static pile.
To reach a steady state or an oscillatory state, a system
of granular media requires an external energy source.
In this paper, we investigate the steady state velocity distributions and
velocity correlations of spatially homogeneous granular media
subject to a volumetric Gaussian white noise forcing.
This system has been studied by several
authors~\cite{vannoije98,dsmc,williams,vannoije99}
as a reference system for the kinetic theory of granular media.
In this system, particles collide inelastically, and execute Brownian motion
between collisions;
the motion is analogous to Brownian dynamics of hard sphere suspension,
but here the kinetic energy is dissipated due to
the inelastic collisions rather than hydrodynamic drag.
This system is far from equilibrium, and the steady state velocity
distribution deviates from the Maxwell-Boltzmann (MB) distribution.
The velocity distribution of this system was first theoretically studied
by van Noije {\it et al.}~\cite{vannoije98},
who obtained the approximate solutions to the inelastic hard sphere
Enskog-Boltzmann equation with a Gaussian white noise forcing.
Their results were tested against the Direct Simulation Monte
Carlo (DSMC) method of the inelastic Enskog-Boltzmann equation~\cite{dsmc},
and a good agreement was found; it confirmed the accuracy of
the approximate analysis of van Noije {\it et al.},
because the validity of the DSMC method relies on the validity
of the inelastic (Enskog-)Boltzmann equation~\cite{bird}.

In the current study, we use a large molecular dynamics (MD) simulation
(using up to $477,500$ particles) of inelastic hard spheres
to investigate the steady state velocity distributions,
and quantitatively examine the accuracy of the inelastic
Enskog-Boltzmann equation model for this system.
Our method is free from the assumptions used in
the inelastic kinetic theory,
and it has an advantage that correlations can develop
but has a disadvantage that it may have a finite simulation size effect.

We also compare our results with theoretical predictions of
Carrillo {\it et al.}~\cite{carrillo}, who obtained the steady
state velocity distribution
by using the pseudo-Maxwell molecule model~\cite{bobylev}.
The Maxwell molecule model is the special case of the inverse
power law model for the inter-particle potential, whose potential has
the form of $V(r) \sim r^{-4}$, where $r$ is the inter-particle distance.
This model has been widely used for analytical studies as a reference
system for more realistic systems, because the model facilitates
calculations involving linearized Boltzmann operator.
The pseudo-Maxwell molecule model is an inelastic analog of
the Maxwell molecule model.

The remainder of the paper is organized as follows.
Section II presents the method of the numerical simulation,
and Section III briefly reviews the theoretical predictions of
the inelastic hard sphere model and the pseudo-Maxwell molecule model.
In Section IV, simulation results of the velocity distributions are
presented, and these are compared with the theoretical predictions
(Sec. IV A).
The simulation results of the velocity correlations are also presented
in this Section (Sec. IV B).
The paper is summarized and concluded in Section V.

\section{numerical simulation}
We simulate an ensemble of inelastic hard sphere particles of diameter
$\sigma$ in a three-dimensional (3d) cubic box of each side $105.3\sigma$,
which are subject to a volumetric Gaussian white noise forcing.
Particles obtain the kinetic energy from the white noise forcing,
execute Brownian motion in between collisions,
and dissipate the kinetic energy through inelastic collisions.
To be consistent with the theoretical studies presented in Section III,
we implement a velocity-independent coefficient of restitution $e$,
and neglect the rotational degrees of freedom.
We use an event-driven MD code originally developed to simulate
the patterns in vertically oscillated granular layers.
Excellent agreement was found between simulations and experiments~\cite{bizon}.
We modify this code to implement the Gaussian white noise forcing.
Brownian dynamics for hard sphere simulation was originated
in the work of Ermak {\it et al.}~\cite{ermak}, and the granular analog
was introduced by Williams {\it et al.}\cite{williams} for their
study of a one-dimensional system.

To implement Gaussian white noise forcing in our MD code,
we start from a stochastic equation of motion for a particle.
The equation of motion is given by

\begin{equation}
\ddot x_i(t) = {\cal F}^{(c)}_i(t) + \Gamma_i(t),
\end{equation}
where the mass of the particle is the unity,
$x_i$ is the $i^{th}$ cartesian component of the position,
${\cal F}^{(c)}_i$ is the $i^{th}$ component of the forcing due to the collisions, 
and $\Gamma_i(t)$ is the $i^{th}$ component of the stochastic forcing.
The stochastic forcing term satisfies

\begin{equation}
<\Gamma_i(t)> = 0,
\end{equation}
which assures that the fluctuation cancels out on the average,
where $<>$ is an ensemble average, and

\begin{equation}
<\Gamma_i(t)\Gamma_j(t')> = 2F\delta_{ij}\delta(t-t'),
\end{equation}
where $F$ is the strength of the correlation,
$\delta_{ij}$ is the Kronecker delta, and $\delta(t)$ is the delta function.
Eq.(3) assures that the collisions well separated in time are
statistically independent.
We assume that all higher-order moments of the random variable
$\Gamma_i(t)$ can be expressed in terms of the second moment,
which is identical to the assumption that $\Gamma_i(t)$ is distributed
according to the Gaussian distribution~\cite{balescu}.
We implement an equation equivalent to Eq.(1) for each particle
with a discrete time interval of a fixed size.
It can be shown that the discrete Langevin equation for the velocity
in between collisions subject to a Gaussian white noise is given by

\begin{equation}
v_i(t+\delta t) = v_i(t) + \sqrt{F}\sqrt{\delta t}G(0,1),
\end{equation}
where $F$ is the same quantity as in Eq.(3),
$\delta t$ is the time interval between the white noise forcing,
and $G(0,1)$ is the unit Gaussian distribution random variable.

In the simulation, $N_k$ particles are randomly chosen at
every $\delta t$ and are kicked in accordance with Eq.(4).
To avoid the development of a mean flow in the system,
particles to be kicked are randomly picked pairwise, and particles of
this pair are kicked in opposite directions with the same speed
to conserve the momentum.
In an event-driven simulation, the random kickings are computationally
expensive discrete events,
because the collision list needs to be updated at each kicking.
For the efficiency of the simulation, the mean kicking frequency needs
to be minimized, or the mean kicking time for each particle
$\bar{\delta t}~(= \delta t \cdot N/N_k)$ needs to be maximized,
where $N$ is the total number of particles.
We checked in the simulation that the velocity distributions
do not change as far as $\bar{\delta t}$ is less than $1/(5\nu_{coll})$,
where $\nu_{coll}$ is the mean collision frequency.
We also checked that the velocity distributions do not depend
on the choice of $N_k$ and $\delta t$ in these cases.
Thus we keep $\bar{\delta t} \sim 1/(10\nu_{coll})$ throughout the simulations.
The volume fraction $\nu$ varies from $\nu_o (= 4.29\%)$ to
$5\nu_o (= 21.4\%)$, which corresponds to $95,495$ to $477,500$
in the number of particles.
Since the collisions are instantaneous, there is only one time scale
determined by the granular temperature, $\sigma/\sqrt{T}$, where $T$
is defined in Eq.(6).
Therefore the temperature only rescales the time,
and the results are independent of $T$.
However, we fix the granular temperature at approximately the same value
throughout the simulations.
We prepare the initial conditions by locating particles in a regular lattice.
We heat them, and wait until the transients decay away,
ensuring that a steady state is reached.
The data are taken periodically with a fixed time interval $\Delta t$.
To assure that each data set is statistically uncorrelated,
$\Delta t$ is chosen larger than $5/\nu_{coll}$ in all cases,
and $\Delta t \sim 10/\nu_{coll}$ in most cases.
We obtain $50$ such data sets for each simulation,
and the error bars appearing in this paper are the standard deviations
of these data sets.
A periodic boundary condition is imposed in all direction.

\section{review of theory}
In this Section, we briefly review the results of the previous theoretical
studies of van Noije {\it et al.}~\cite{vannoije98} and of Carrillo
{\it et al.}~\cite{carrillo}. Both studies are based on the mean field-type
inelastic Enskog-Boltzmann equation.

Following standard procedures of the kinetic theory,
it can be shown that the inelastic Enskog-Boltzmann equation for
a system of spatially homogeneous granular particles
subject to a white noise forcing is given by

\begin{equation}
{\partial f \over \partial t} = Q(f,f) + L_{FP}f,
\end{equation}
where $f = f({\bf c},t)$ is the single particle distribution function, 
$Q(f,f)$ is the collision operator,
${\bf c} = {\bf v}/\sqrt{2T(t)}$ is the velocity scaled by
the characteristic velocity, and the mass is the unity.
$T(t)$ is the granular temperature, defined as the variance of the
velocity distribution per degree of freedom. For a 3d system,

\begin{equation}
T(t) = {1 \over 3}<|{\bf v}(t) - <{\bf v}(t)>|^2>.
\end{equation}
The Fokker-Planck operator $L_{FP}$ is a diffusion operator in
the velocity space, which is given by

\begin{equation}
L_{FP} = F\nabla^2_{c},
\end{equation}
where $F$ is the strength of the correlation of
Gaussian white noise stochastic forcing (defined in Eq.(3)),
and $\nabla^2_{c}$ is the Laplacian in the velocity space.
The collision operator, $Q(f,f)$, is chosen depending on the inter-particle
potential model, and the most obvious one for the granular particles is
the inelastic hard sphere collision operator.

Van Noije {\it et al.}~\cite{vannoije98} obtained the steady state
solutions to Eq.(5) using the inelastic hard sphere collision operator,
and Carrillo {\it et al.}~\cite{carrillo} obtained the steady state 
solutions using the pseudo-Maxwell collision operator.
In both analyses, the Sonine polynomial expansion method was used
to construct the solution.
Sonine polynomials are associated Laguerre polynomials and
have been used to construct solutions to the Boltzmann equation
since Burnett~\cite{burnett} introduced them in the study of nonuniform gases.
They are exact eigenfunctions of the linearized Boltzmann
equation for Maxwell molecules.
This expansion also leads to rapidly converging solutions of the linearized
Boltzmann equation for other short-range repulsive potentials\cite{chapman}.
The Sonine polynomials of lower parameter $1/2$ is defined by 

\begin{equation}
S^{(n)}_{1/2}(x) = \sum_{p=0}^n {({1\over 2}+n)! \over ({1\over 2}+p)!(n-p)!p!}(-x)^p.
\end{equation}
In the current study, $c^2$ is the argument of Sonine polynomials.
The orthogonality relation with this argument is

\begin{equation}
\int_{0}^{\infty}c^2e^{-c^2}S^{(n)}_{1/2}(c^2) S^{(m)}_{1/2}(c^2) dc = {1\over 2}\delta_{nm}{({1 \over 2}+n)!\over n!}.
\end{equation}
The first three Sonine polynomials are given by

\begin{eqnarray*}
S^{(0)}_{1/2}(c^2) &=& 1, \\
S^{(1)}_{1/2}(c^2) &=& {3 \over 2} - c^2, \\
S^{(2)}_{1/2}(c^2) &=& {15 \over 8} - {5 \over 2}c^2 + {1 \over 2}c^4.
\end{eqnarray*}

For the inelastic hard sphere model, van Noije {\it et al.}~\cite{vannoije98}
obtained the steady state solutions for 2d and 3d cases,
by using the moment method~\cite{goldshtein}
which was used in the analysis of the velocity distributions of
homogeneously cooling of a freely evolving system.
They expanded the solution in Sonine polynomials and neglected terms of
higher order than the first nonvanishing correction to
the MB distribution.
The steady state solution for a 3d case, $f^s(c)$, is obtained as

\begin{equation}
f^s_{(HS)}(c) = f_{MB}(c)\lbrack 1 + a_2^{HS}S^{(2)}_{1/2}(c^2)\rbrack,
\end{equation}
where 

\begin{equation}
f_{MB}(c) = {4 \over \sqrt{\pi}}c^2 \exp(-c^2)
\end{equation}
is the MB distribution (multiplied by an integration factor),
and

\begin{equation}
a_2^{HS}(d,e) = {16(1-e^2)(1-2e^2)\over73+56d-24ed-105e+30(1-e)e^2},
\end{equation}
where $d$ is the dimensionality of the system ($d = 3$ in the current study),
and $HS$ stands for inelastic hard spheres.

For the pseudo-Maxwell molecule model, Carrillo {\it et al.}~\cite{carrillo}
obtained the steady state solution to Eq.(5) by doubly expanding
the solution in the energy dissipation rate, $\epsilon = (1-e^2)/4$,
and in Sonine polynomials.
The deviation from the MB distribution was obtained
up to the order of $\epsilon^4$~\cite{carrillo2}.
The solution with the first nonvanishing correction to the MB
distribution is of the order of $\epsilon^2$, which is given by

\begin{equation}
f^s_{(MM)}(c) = f_{MB}(c)\lbrack 1 + \epsilon^2 \cdot 4S^{(2)}_{1/2}(c^2) \rbrack,
\end{equation}
where $MM$ stands for pseudo-Maxwell molecules.
The first order deviations from the MB distribution
in the two models have the same basis function $S^{(2)}_{1/2}(c^2)$,
the second Sonine polynomial.
We rewrite the normalized deviations from the MB in the two models as follows,

\begin{eqnarray}
g^{HS}(e,c) &=& {\Delta f^{HS}(e,c) \over f_{MB}(c)} = a_2^{HS}(e) S^{(2)}_{1/2}(c^2),\\
g^{MM}(e,c) &=& {\Delta f^{MM}(e,c) \over f_{MB}(c)} = a_2^{MM}(e) S^{(2)}_{1/2}(c^2),
\end{eqnarray}
where
\begin{equation}
a_2^{MM}(e) = (1 - e^2)^2/4 = 4\epsilon^2.
\end{equation}
The coefficients of the second Sonine polynomial for the two models,
$a_2^{HS}(e)$ and $a_2^{MM}(e)$, are compared in Fig. 1.

To summarize, the predictions of the inelastic hard sphere model and those of
the pseudo-Maxwell molecule model are different mainly in two ways: 

(1) There is a crossover from positive to negative values in $a_2^{HS}(e)$
as $e$ increases, and it is negative for $1/\sqrt{2} < e < 1.0$,
while $a_2^{MM}(e)$ is positive definite, $4\epsilon^2$.
The high energy tail is always overpopulated in the pseudo-Maxwell
molecule model, while it is underpopulated for $e > 1/\sqrt{2}$
in the inelastic hard sphere model, when the series is truncated
at this order.

(2) The deviation from the MB distribution in
the pseudo-Maxwell molecule model is quantitatively much larger than
that in the inelastic hard sphere model.
Note that the analysis of the pseudo-Maxwell molecule model
uses a small inelasticity assumption, while the analysis of
the inelastic hard sphere model has no such assumption.

\section{simulation results}
\subsection{velocity distributions}
We measure the velocity of each particle periodically in time
with a fixed time interval of $\Delta t \sim 10/\nu_{coll}$,
which is chosen to assure that each data set is statistically uncorrelated.
The measured velocities are binned, and the bin size is $\Delta c = 0.1$.
The velocity distributions for two coefficients of restitution,
$e = 0.1$ and $e = 0.9$, are obtained in the simulation (Fig. 2).
As in the inelastic hard sphere theory, high velocity tail is overpopulated
for $e = 0.1$, and slightly underpopulated for $e = 0.9$,
compared to the MB distribution.

We define the normalized deviation from the MB distribution,
$g(c)$, which is obtained in the simulation,

\begin{equation}
f_{(MD)}^s(c) = f_{MB}(c)\lbrack 1 + g(c) \rbrack,
\end{equation}
where the subscript $(MD)$ means this is obtained in the MD simulation.
Assuming $g(c)$ is a smooth function in the scale of
$\Delta c~(\ll 1)$, $g(c)$ can be approximated as

\begin{equation}
g(c_o+{\Delta c\over 2}) \approx {\int_{c_o}^{c_o+\Delta c} f_{(MD)}^s(c)dc \over \int_{c_o}^{c_o+\Delta c}{4\over\sqrt{\pi}}e^{-c^2}c^2 dc}~-~1.
\end{equation}
The numerator in Eq.(18) is the total number of particles in the bin,
which is a number counted in the simulation,
and the denominator can be evaluated using the error function table.
$g(c)$ calculated using Eq.(18) for the data in Fig. 2 is shown in Fig. 3.
For direct comparison between our simulation results and
the above theoretical predictions,
we calculate the coefficient of the second Sonine polynomial,
$a_2^{MD}$, from our measurements.
As in the theoretical analyses,
we assume $f_{(MD)}^s(c)$ is expanded in Sonine polynomials,

\begin{equation}
f_{(MD)}^s(c) = f_{MB}(c)\lbrack 1 + \sum_{k=2}^{\infty}a_k^{MD}S^{(k)}_{1/2}(c^2)\rbrack,
\end{equation}
where $a_1^{MD}$ is not included, because it is identically zero in theory.
We checked in the simulation that $a_1^{MD}$ is less than $10^{-4}$ in all cases.

In the simulation, we can use either of the following two formulae to
calculate $a_k^{MD}$, the coefficient of the $k^{th}$ Sonine polynomial.
Firstly, $a_k^{MD}$'s can be obtained from the following numerical integration,
using the orthogonality relation for Sonine polynomials,

\begin{equation}
a_k^{MD} = {2^k k!\over (2k+1)!!} \int_{0}^{\infty} f_{(MD)}^{s}(c) S_{1/2}^{(k)}(c^2)dc.
\end{equation}
Secondly, we can make use of the recurrence relation of $a_k^{MD}$'s.
Starting from the definition of Sonine polynomials, Eq.(8),
it can be shown that for $k > 2$, $a_k^{MD}$ satisfies

\begin{equation}
a_k^{MD} = {(-1)^k 2^k\over (2k+1)!!}<c^{2k}> + (-1)^{k+1} + \sum_{p=1}^{k-2} (-1)^{p+1} {k \choose p} a_{k-p}^{MD},
\end{equation}
where $<c^{2k}>$ is the $2k^{th}$ moment.
It is straightforward to numerically evaluate the integration in both formulae.
The above two relations, Eq.(20) and Eq.(21), are mathematically identical.
However, the errors of the lower $k$ coefficients are accumulated on higher
$k$ coefficients in Eq.(21), while each $a_k^{MD}$ is determined independently
in Eq.(20).
The results obtained using Eq.(20) and Eq.(21) are nearly the same
for small $k$, and they deviate more for larger $k$.
$a_2^{MD}$ is obtained using Eq.(20), which are compared with
the theoretical predictions Eq.(12), $a_2^{HS}$, and Eq.(16), $a_2^{MM}$
(Fig. 4).
The simulation results deviate more from the predictions of
the inelastic hard sphere kinetic theory as the system becomes
more inelastic, and $a_2^{MD}$ has a crossover between the positive and
the negative values at around $e \sim 0.8$, while it was predicted to
occur at $1/\sqrt{2}$ by the inelastic hard sphere theory. 
The pseudo-Maxwell molecule model does not predict a crossover
behavior of $a_2^{MM}$, and the predictions of this model deviate
quantitatively from the kinetic theory and simulations of
the inelastic hard sphere model.

In the simulation, all $a_k^{MD}$'s can be obtained by using Eq.(20).
For larger $k$, the weight of the high velocity data exhibiting
strong fluctuations become larger, and so does the uncertainty of $a_k^{MD}$.
We calculate up to $a_5^{MD}$ for various coefficients of restitution (Fig. 5).
For nearly elastic case, for instance $e = 0.9$, the series converges
rapidly as in many cases of elastic hard spheres.
However, as the system becomes more inelastic, the series converges
more slowly. For $e = 0.1$, $a_4^{MD}$ is about $30\%$ of $a_2^{MD}$. 
It is shown that how the first few nonvanishing coefficients of
the Sonine polynomial series obtained in the simulation
affect the fitting (Fig. 6).
For $e = 0.1$, $a_2^{MD}$ fits the measured data well up to
$v \sim 3\sqrt{2T} \sim 4\sqrt{T}$, but the high energy tail is
fitted well only when the series are kept up to $a_4^{MD}$.
For $e = 0.9$, it shows the same tendency.
When we fit the data only with $a_2^{MD}$ in this case, $f^s_{(MD)}(c)$
becomes negative for $c > 4.1$ and becomes negative infinity
as $c$ goes to infinity.
Such nonphysical behavior disappears when we include $a_3^{MD}$.

In the theoretical predictions, the steady state velocity distribution
and $a_2^{HS}$ do not depend on the density, and it is a function of
the coefficient of restitution and the dimensionality only.
However, we find that it also depends on the density (Fig. 7).
The value of $a_2^{MD}$ decreases as the system becomes more dilute.
For the smallest volume fraction we used, $\nu_o$,
the value of $a_2^{MD}$ deviates by only few percent from the predictions
of the inelastic hard sphere theory.
We plot $a_4^{MD}$'s for various densities in Fig. 8 ($a_3^{MD}$'s are
very small, as shown in Fig. 5).
It follows the same tendency as $a_2^{MD}$ does;
as the system becomes denser, $a_4^{MD}$ gets larger.

Finally, we examine the asymptotic behavior of the high energy tails
of the velocity distribution functions.
Van Noije {\it et al.}~\cite{vannoije98} found for this system
a high energy tail shows an asymptotic behavior
$\sim \exp(-{\cal A}'c^{3/2})$.
We measure the velocity distribution function ${\tilde f}^s_{(MD)}(c)$, 
which is defined as

\begin{eqnarray}
1 &=& \int {\tilde f}^s_{(MD)}(c) d{\bf c}\cr
&=& \int {\tilde f}^s_{(MD)}(c)4\pi c^2 dc\cr
&=& \int f^s_{(MD)}(c) dc.
\end{eqnarray}
To investigate the power of the argument of the exponential function,
we renormalize ${\tilde f}^s_{(MD)}(c)$ with its maximum value.
We observe the crossover behavior from $\sim \exp(-{\cal A}c^2)$
to $\sim \exp(-{\cal A}'c^{3/2})$ as $c$ increases, for $e \leq 0.5$ (Fig. 9).

\subsection{Velocity correlations}
Two major approximations imposed in the kinetic theory discussed
in Section III are the mean field approximation and the truncation of
the hierarchy by introducing the molecular chaos assumption.
In this Section, we examine the validity of each of the above approximations
for this system by quantitatively investigating the parallel velocity
correlations, in long-range and short-range respectively.
It is known that a system of granular media exhibits strong spatial
correlations~\cite{williams,correlation,vannoije99},
such as velocity correlations,
but there is still the lack of a quantitative study of a 3d system.
We suggest that the deviation of $a_2^{MD}$ from $a_2^{HS}$ originates
from the failure of the above approximations.

We define the parallel velocity correlation function as

\begin{equation}
<c_{1,||}c_{2,||}> = \nonumber\\
{1 \over N} \int <{\bf c(r+r')}\cdot {\bf \hat{r}}\rho({\bf r}+{\bf r'}){\bf c(r')}\cdot {\bf \hat{r}}\rho({\bf r'})>d{\bf r'},
\end{equation}
where ${\bf \hat{r}} = {\bf r}/|{\bf r}|$,
and $\rho({\bf r})$ is the local particle density,

\begin{equation}
\rho({\bf r}) = \sum_{i=1}^{N} \delta({\bf r} - {\bf r}_{i}).
\end{equation}
We approximately evaluate this quantity using the following formula,

\begin{equation}
<c_{1,||}c_{2,||}> = \sum {{\bf c}_{1,||}{\bf c}_{2,||} \over N_r},
\end{equation}
where the parallel direction is along the line of centers of
the particle pair under consideration, $N_r$ is the number of particles
in a shell of thickness $\delta r$ and inner radius $r$,
and the summation is done over $N_r$.
We use $\delta r = 0.1053\sigma$.

\subsubsection{Long-range correlations}
The parallel velocity correlations are obtained by averaging
over $50$ statistically uncorrelated data sets.
We calculate those for various coefficients of restitution
as a function of dimensionless distance $r/\sigma$ (Fig. 10),
and for various densities (Fig. 11).
The data are shown only for $r/\sigma < 30$,
because they are subject to the finite system size effect
for larger $r/\sigma$.
The parallel velocity correlations in our simulation are consistent with
slow algebraic decay over a decade, $\sim r^{-(1+\delta)}$,
where $0.2 < \delta < 0.3$.
This behavior is close to the theoretical prediction
of van Noije {\it et al.}~\cite{vannoije99},
who predicted the $r^{-1}$ power law from the mode coupling theory.
The correlations in our simulation get stronger for more inelastic or
more dense system, which implies that the mean field approximation is
reliable only for nearly elastic and dilute cases.

\subsubsection{Short-range correlations}
In this section, we investigate the velocity correlations at contact before
the collision, to examine the validity of the molecular chaos assumption.
A non-zero value of these correlations is the signature of the failure
of the molecular chaos assumption;
the velocities are more ``parallelized'' after the collision,
since only the normal component of relative velocity of colliding particles
are reduced at the collision, which means that
if the velocities are already parallelized before the collision,
it would indicate that the colliding pair have ``memory'' on the collisions
in the past, and they are correlated.
We find that for high inelasticity and density, the pre-collisional
parallel velocity correlation value reaches up to $\sim 15\%$
of the temperature.
Even for dilute or nearly elastic cases, these are not negligibly small.

We calculate the velocity correlations of pre- and post-collisional
states by evaluating Eq.(25) for approaching
(${\bf r}_{12}\cdot {\bf c}_{12} < 0$) and separating particles
(${\bf r}_{12}\cdot {\bf c}_{12} > 0$) respectively,
where ${\bf x}_{12} = {\bf x}_{1} - {\bf x}_{2}$
for ${\bf x} = {\bf r}$ or ${\bf c}$.
We calculate the pre-collisional state of short-range
($1 < r/\sigma < 2$) parallel velocity correlations (Fig. 12),
and their post-collisional state (Fig. 13),
for various coefficients of restitution.
The values of the pre-collisional correlations are not negligible
compared to the temperature of the system.
Post-collisional correlations are more than twice as large as
pre-collisional correlations.

The maximum values of the velocity correlations in Fig. 12
and in Fig. 13 are not the values at the contact, $r = \sigma$,
because of the finite size of the bins in the measurements;
instead, those are values at $r/\sigma = 1.053$.
The values at $r = \sigma$ are estimated by extrapolating
the data in the interval $1 < r/\sigma < 2$ using the least square fit
with fifth-order polynomials.
We estimate the velocity correlations at contact, $r = \sigma$,
as a function of the coefficient of restitution (Fig. 14),
and as a function of the density (Fig. 15).
Since the velocity correlation varies rapidly as $r/\sigma$ decreases
to $1$, these estimations may be regarded only as approximate lower bounds.
The velocity correlations at contact in this system show almost linear
behavior both in density and the coefficient of restitution.

\section{Discussion}
We have investigated the velocity distributions and
parallel velocity correlations of 3d homogeneously heated granular media
for various densities and inelasticities,
using an inelastic hard sphere MD simulation.
The deviations from the MB distribution in our
simulations qualitatively agree with the results of
the mean field-type inelastic hard sphere kinetic theory~\cite{vannoije98},
but we found that there is systematic quantitative difference.

We observed the high energy tails are consistent with
$\sim \exp(-{\cal A}'c^{3/2})$ for $e \leq 0.5$, but not for others.
Since the elastic case ($e = 1.0$) has no crossover from
$\sim \exp(-{\cal A}c^2)$ to $\sim \exp(-{\cal A}'c^{3/2})$,
we expect that this crossover behavior may occur at higher
velocities as $e$ approaches to $1.0$, if it occurs.
However, we were not able to check whether the crossover
occurs for $e > 0.5$ or never occurs, because our system is finite.
It is interesting to note that the same behavior was experimentally
observed in a system with different forcing mechanism~\cite{losert}.

We found that the steady state velocity distributions in the simulation
depend on the density as well as the coefficient of restitution,
while they depend only on the latter in the theory.
The discrepancy between our simulation results and the theoretical
predictions
increases as the system becomes more inelastic or more dense,
and the quantitative disagreement reaches up to $\sim 40\%$.
This behavior is consistent with the results of
van Noije {\it et al.}~\cite{vannoije99,trizac}, who found that the collision
frequency measured in a 2d MD simulation deviates more from the predictions
of the inelastic Enskog-Boltzmann equation as the system becomes more
inelastic.
We suggest that the disagreement originates from the failure of two
major approximations in the theory, the mean field approximation
and the molecular chaos assumption.
To examine the accuracy and the limitation of these approximations,
we quantitatively investigated the parallel velocity correlations
of this system.

We found that the long-range parallel velocity correlations are
consistent with a slow algebraic decay, $\sim r^{-(1+\delta)}$,
where $0.2 < \delta < 0.3$.
This result is close to the theoretical predictions of
van Noije {\it et al.}~\cite{vannoije99}, who renormalized various
quantities such as the collision frequency using the mode coupling
theory and predicted $r^{-1}$ power law for velocity correlations
in the system we studied.
Because of these strong correlations, the mean field approximation is
not a good one unless the system is nearly elastic or very dilute.

We also found that the velocity correlations at contact
before the collision are not negligible.
We measured the short-range velocity correlations of pre-
and post-collisional states separately to examine the validity
of the molecular chaos assumption.
The pre-collisional correlations at contact are about a half of
the post-collisional correlations.
The correlations at contact are almost linearly proportional to both
the density and the coefficient of restitution, which is consistent with
the recent results of Soto {\it et al.}~\cite{soto},
who studied the velocity correlations of a 2d
homogeneously cooling granular media in nearly elastic regime.

We also examined the convergence of the Sonine polynomial expansion
technique used in the inelastic kinetic theory,
and found that the series converges more slowly
as the system becomes more inelastic.

Finally, we compared the steady state velocity distributions
in the simulations with the theoretical predictions of
Carrillo {\it et al.}~\cite{carrillo},
who studied the current system using the pseudo-Maxwell molecule model.
We found that the velocity distribution function predicted by
this model differ qualitatively from
those predicted by the inelastic hard sphere model.

\section*{acknowledgments}
The authors thank J. A. Carrillo, C. Bizon, A. Santos, I. Gamba,
Daniel I. Goldman, W. D. McCormick and Harry L. Swinney for helpful discussion.
This work was supported by the Engineering Research Program of the
Office of Basic Energy Sciences of the Department of Energy
(Grant DE-FG03-93ER14312).

\setcounter{equation}{0}
\section*{appendix}
\renewcommand\theequation{\thesubsection\arabic{equation}}
\subsection{Derivation of Eq. (18)}
The deviation from the MB distribution, $g(c)$,
is defined as in Eq.(17),

\begin{equation}
f^s_{(MD)}(c) = f_{MB}(c)\lbrack 1 + g(c) \rbrack.
\end{equation}
In the simulation, the number of particles in each bin is measured;

\begin{equation}
\int_{c_o}^{c_o+\Delta c} f^s_{(MD)}(c)dc = \int_{c_o}^{c_o+\Delta c}{4\over\sqrt{\pi}}e^{-c^2}c^2\lbrack 1+g(c)\rbrack dc,
\end{equation}
where the bin size $\Delta c$ is assumed to be very small.
It is possible to numerically solve for $g(c)$ from Eq.(A2), however, 
we get an approximate expression for $g(c)$ by assuming $g(c)$ is smooth
in the scal of $\Delta c$; 

\begin{equation}
\int_{c_o}^{c_o+\Delta c}e^{-c^2}c^2g(c) dc \approx g(c_o+{\Delta c\over 2}) \int_{c_o}^{c_o+\Delta c}e^{-c^2}c^2 dc.
\end{equation}
Eq.(A2) can be read

\begin{equation}
g(c_o+{\Delta c\over 2}) \approx {\int_{c_o}^{c_o+\Delta c} f^s_{(MD)}(c)dc \over \int_{c_o}^{c_o+\Delta c}{4\over\sqrt{\pi}}e^{-c^2}c^2 dc}~-~1.
\end{equation}

\setcounter{equation}{0}
\subsection{Derivation of Eq. (21)}
Starting from the definition of Sonine polynomials,

\begin{equation}
S^{(n)}_{1/2}(x) = \sum_{p=0}^n {({1\over 2}+n)! \over ({1\over 2}+p)!(n-p)!p!}(-x)^p,
\end{equation}
it can be shown that 

\begin{equation}
c^{2k} = \sum_{p=0}^k {(-1)^{k+p}({1\over 2}+k)!k! \over ({1\over 2}+k-p)!p!}S^{(k-p)}_{1/2}(c^2).
\end{equation}
We assume the following Sonine polynomial expansion of
the distribution function,

\begin{equation}
f^s_{(MD)}(c) = f_{MB}(c)\lbrack 1 + \sum_{p=2}^{\infty}a_p^{MD}S^{(p)}_{1/2}(c^2)\rbrack.
\end{equation}
Using Eq.(B2) and Eq.(B3), $2k^{th}$ moment reads

\begin{equation}
<c^{2k}> = {(2k+1)!!\over 2^k}\lbrack 1 + \sum_{p=0}^{k-2} (-1)^{k+p} {k \choose p} a_{k-p}^{MD} \rbrack,
\end{equation}
or

\begin{equation}
a_k^{MD} = {(-1)^k 2^k\over (2k+1)!!}<c^{2k}> + (-1)^{k+1} + \sum_{p=1}^{k-2} (-1)^{p+1} {k \choose p} a_{k-p}^{MD},
\end{equation}
where $k > 2$.

\end{multicols}

\pagebreak

\begin{figure}
\epsfxsize=.98\columnwidth
\centerline{\epsffile{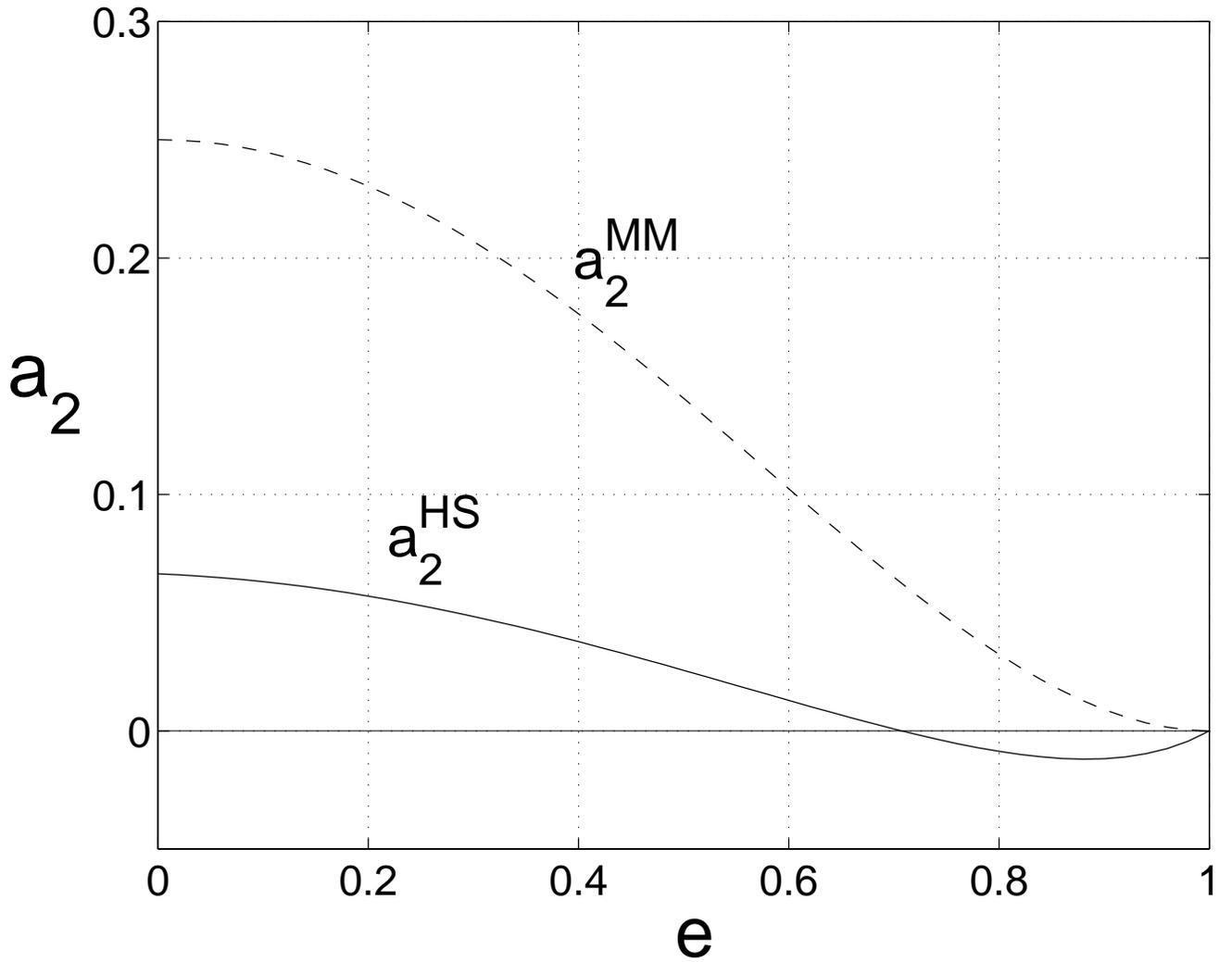}}
\smallskip
\caption{
The deviation from the MB distribution predicted
by the theory, $a_2$, as a function of the coefficient of restitution $e$.
$a_2^{HS}(e)$ is the coefficient of the second Sonine polynomial
from the inelastic hard sphere model, and $a_2^{MM}(e)$ is that from
the pseudo-Maxwell molecule model.
}
\end{figure}
\pagebreak

\begin{figure}
\epsfxsize=.98\columnwidth
\centerline{\epsffile{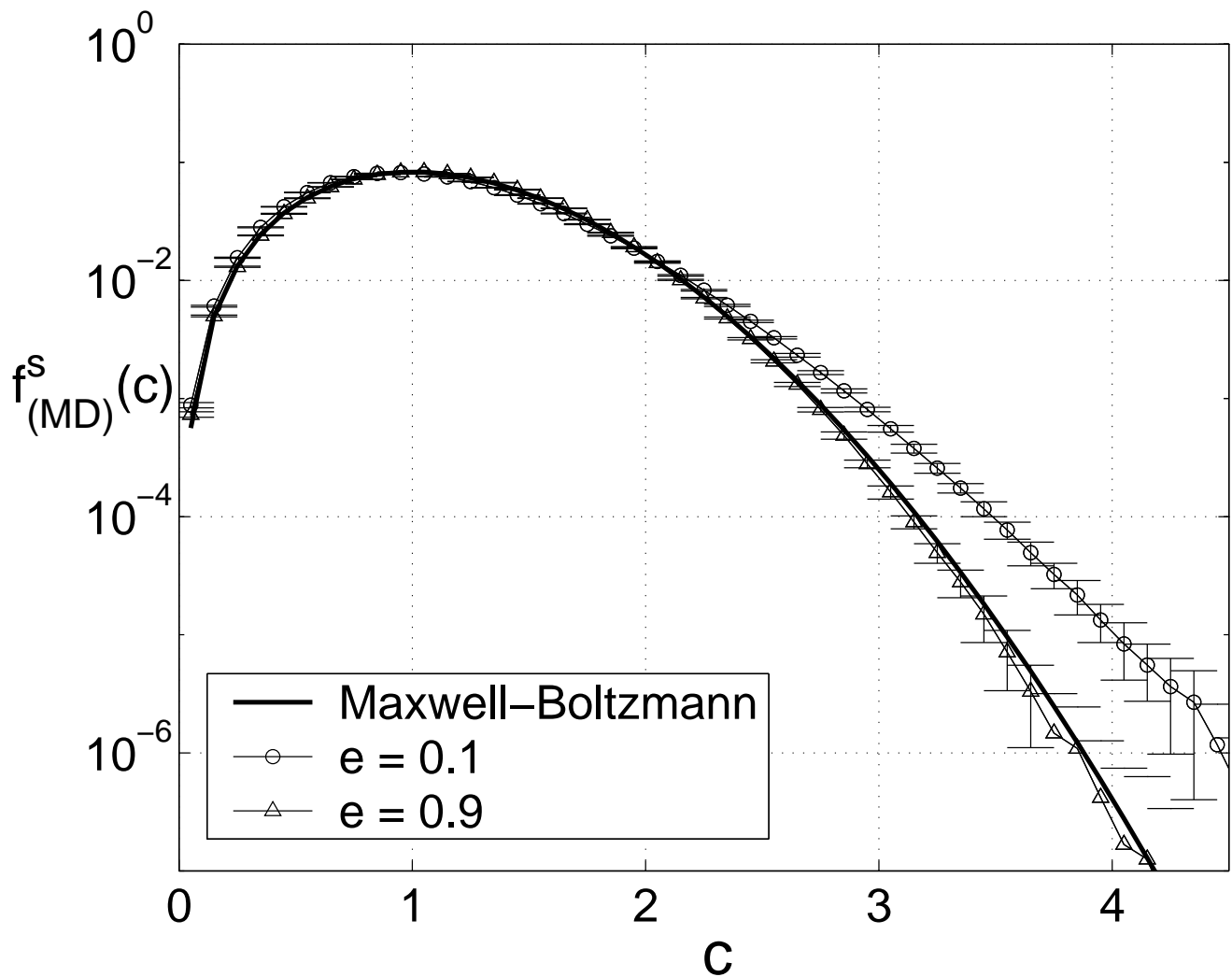}}
\smallskip
\caption{
The steady state velocity distributions, $f^s_{(MD)}(c)$,
obtained from the simulation for two coefficients of restitution,
$e = 0.1$ and $e = 0.9$. The volume fraction is $5\nu_o$.
The thick solid line is the MB distribution function.
}
\end{figure}
\pagebreak

\begin{figure}
\epsfxsize=.98\columnwidth
\centerline{\epsffile{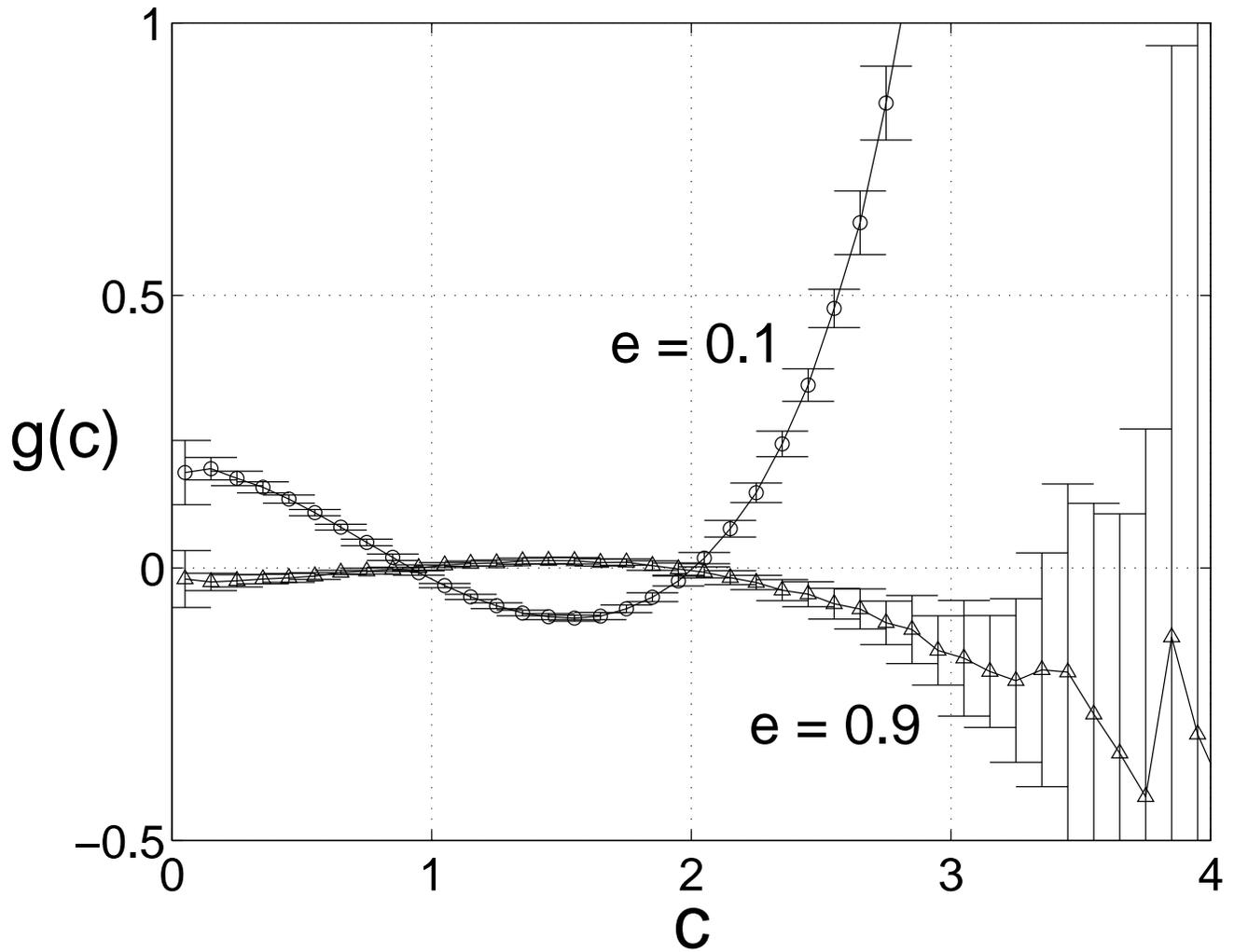}}
\smallskip
\caption{
The normalized deviation from the MB distribution, $g(c)$,
obtained from the simulation 
for two coefficients of restitution, $e = 0.1$ and $e = 0.9$.
The volume fraction is $5\nu_o$.
The error is large for $c > 3$, because it involves division by
a very small number.
}
\end{figure}
\pagebreak

\begin{figure}
\epsfxsize=.98\columnwidth
\centerline{\epsffile{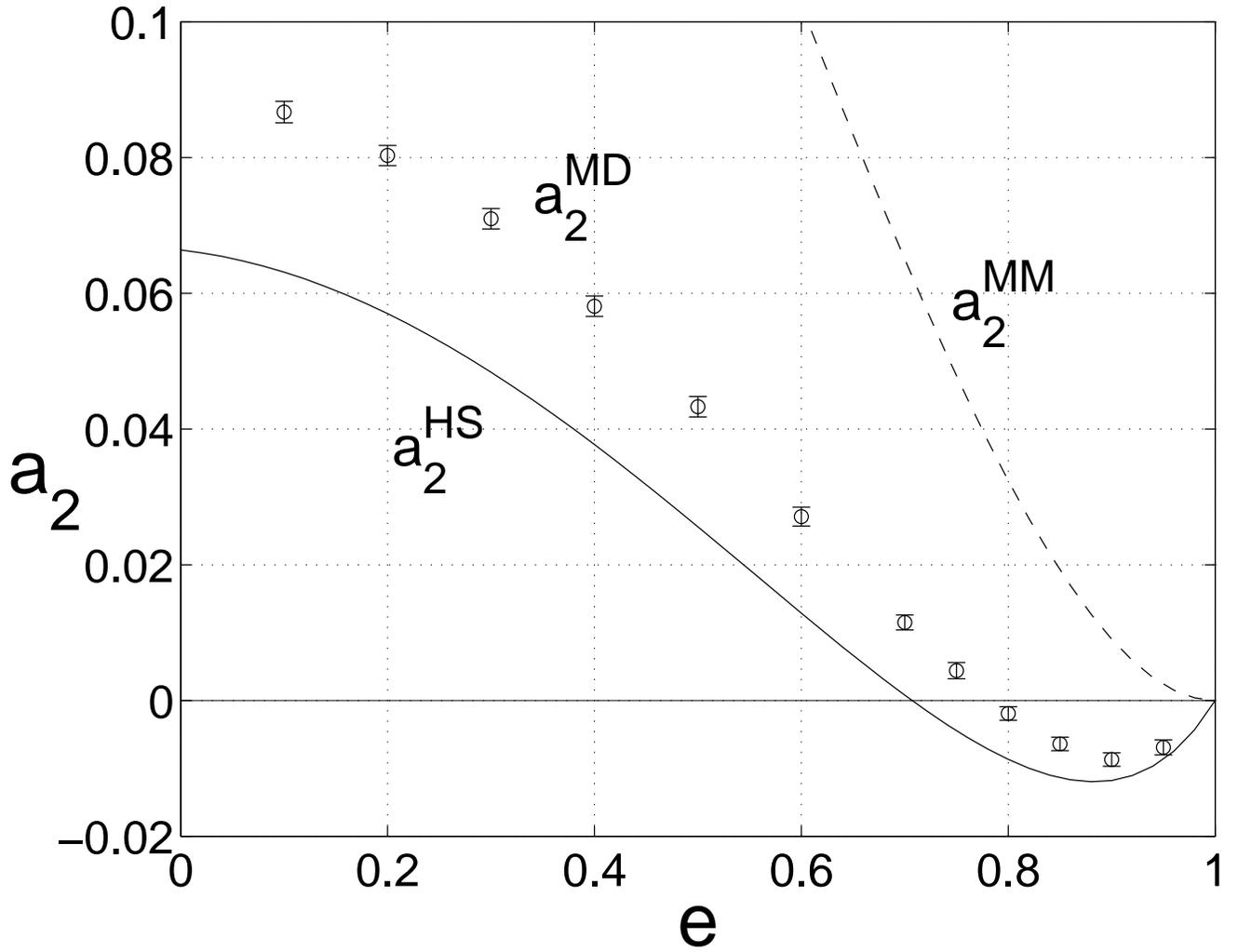}}
\smallskip
\caption{
A comparison of $a_2$'s from the three models, the kinetic theory of
the inelastic hard sphere model (solid curve), the kinetic theory of
the pseudo-Maxwell molecule mode (dashed curve),
and the MD simulation (open circles).
The volume fraction is $5\nu_o$ for the simulation.
}
\end{figure}
\pagebreak

\begin{figure}
\epsfxsize=.98\columnwidth
\centerline{\epsffile{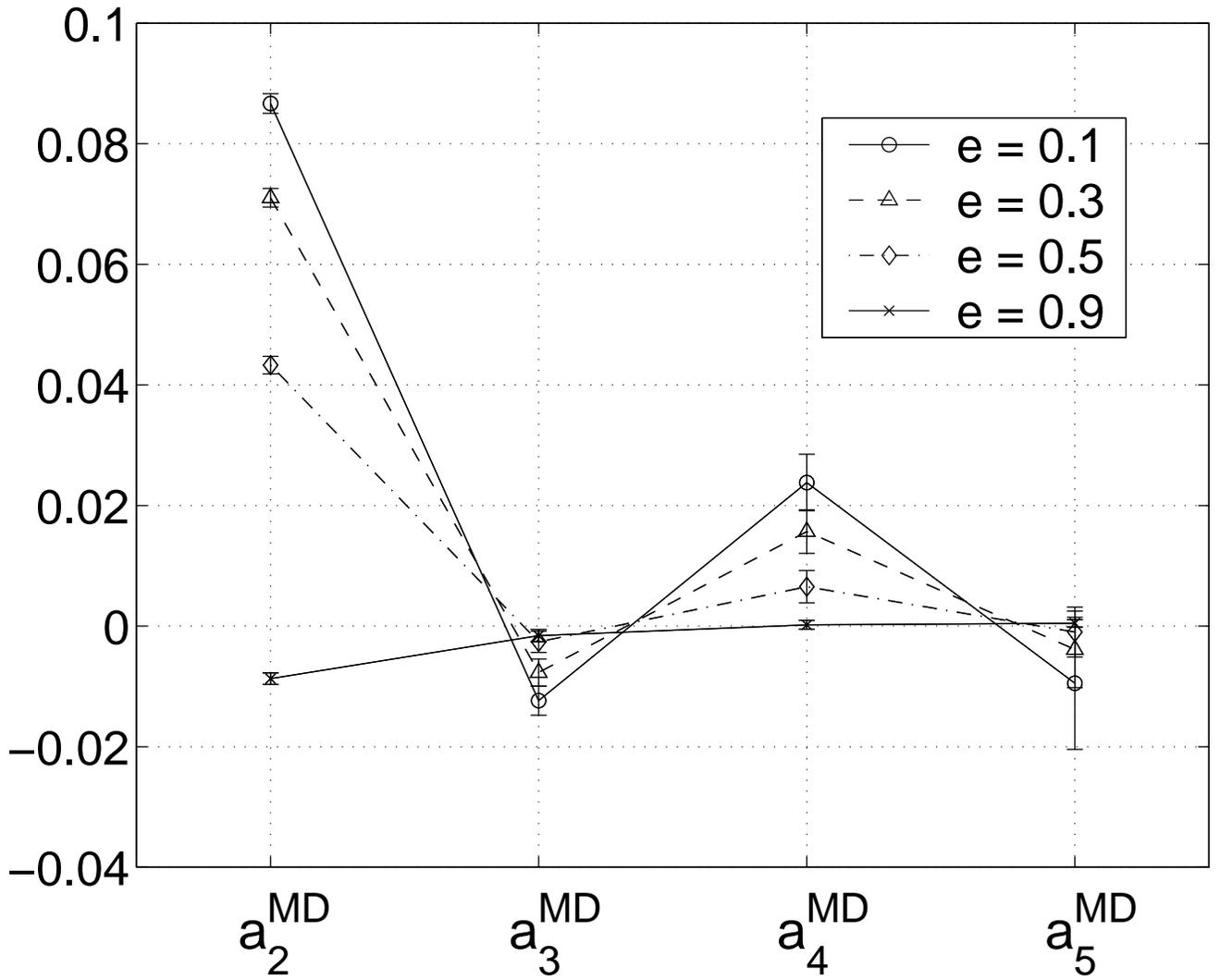}}
\smallskip
\caption{
The Sonine polynomial series converges slower as $e$ decreases.
The first four nonvanishing coefficients, from $a_2^{MD}$ to $a_5^{MD}$,
are calculated, which are obtained from the simulation for
four coefficients of restitution.
The volume fraction is $5\nu_o$.
}
\end{figure}
\pagebreak

\begin{figure}
\epsfxsize=.98\columnwidth
\centerline{\epsfxsize=.49\columnwidth \epsffile{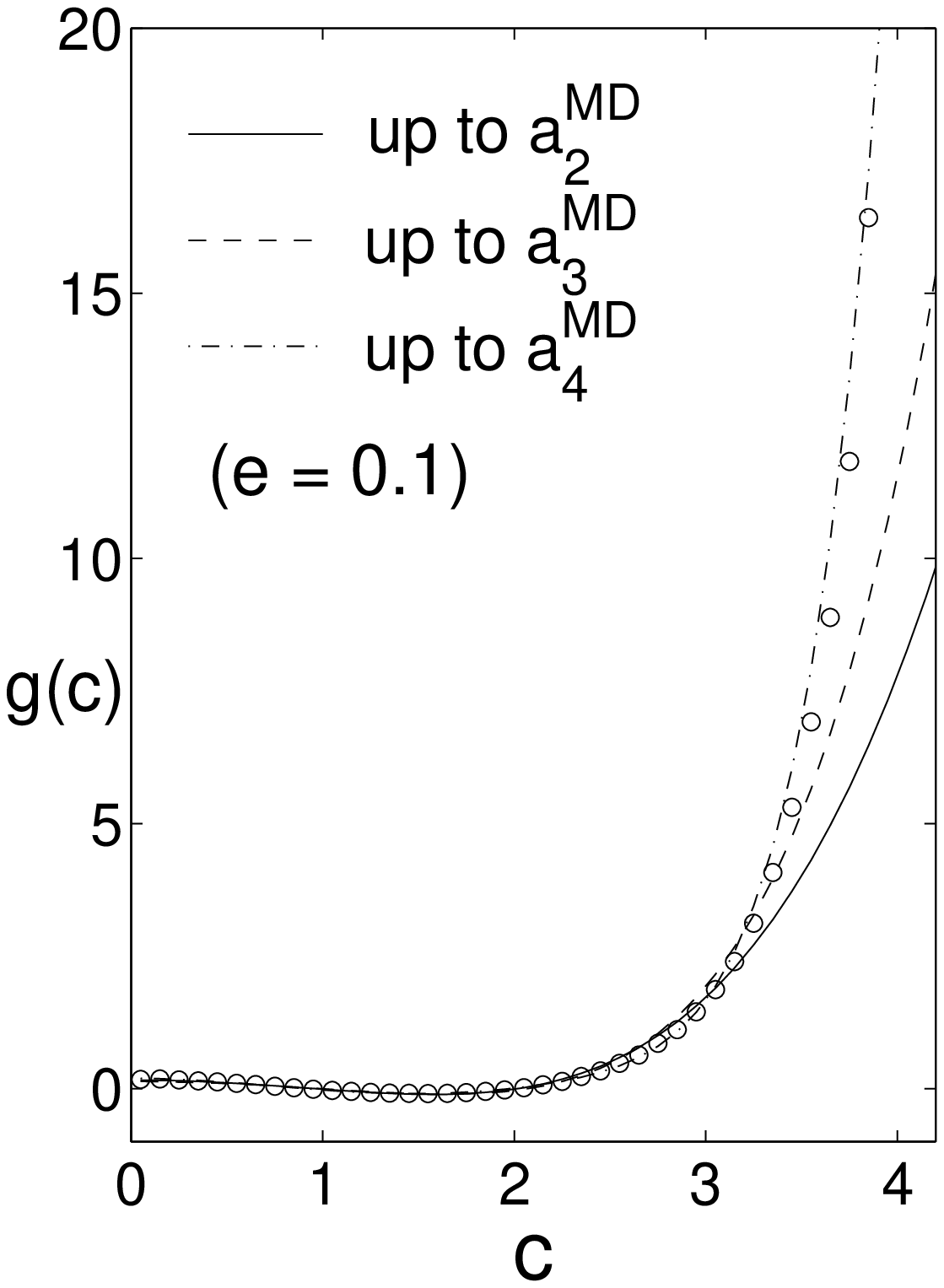}
\hskip .01\columnwidth
\epsfxsize=.49\columnwidth \epsffile{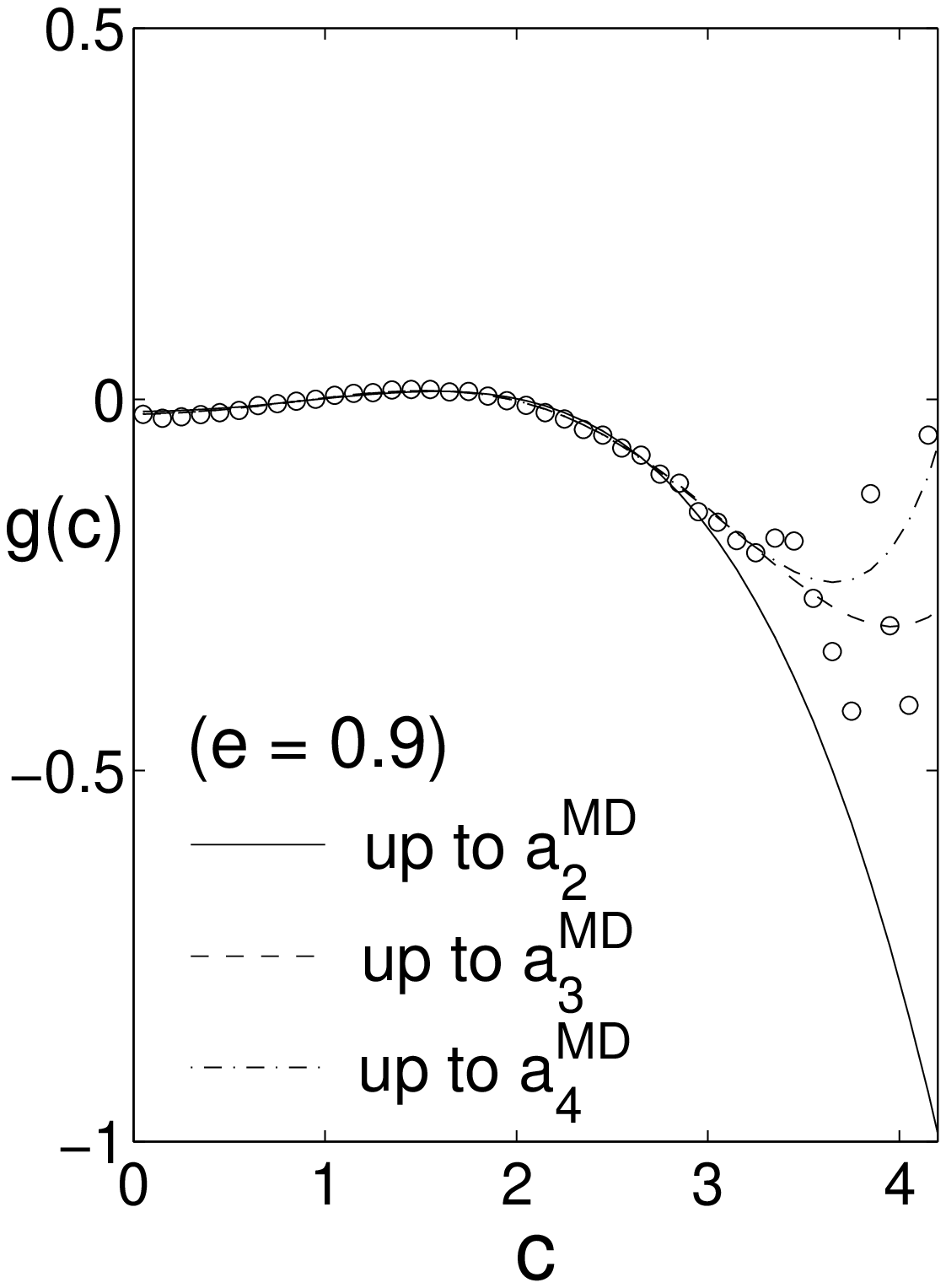}}
\smallskip
\caption{
The normalized deviations from the MB distribution,
$g(c)$, obtained in the simulation (open circles), are compared with
the fitted curves using the coefficients of the Sonine polynomial expansion
calculated using the simulation data,
which are shown for two coefficients of restitution,
$e = 0.1$, and $e = 0.9$.
The terms are successively added to the Sonine polynomial expansion
up to $a_4^{MD}$.
The volume fraction is $5\nu_o$ for both, and the error bars for the data
are not included for better comparison with fitting curves.
}
\end{figure}
\pagebreak

\begin{figure}
\epsfxsize=.98\columnwidth
\centerline{\epsffile{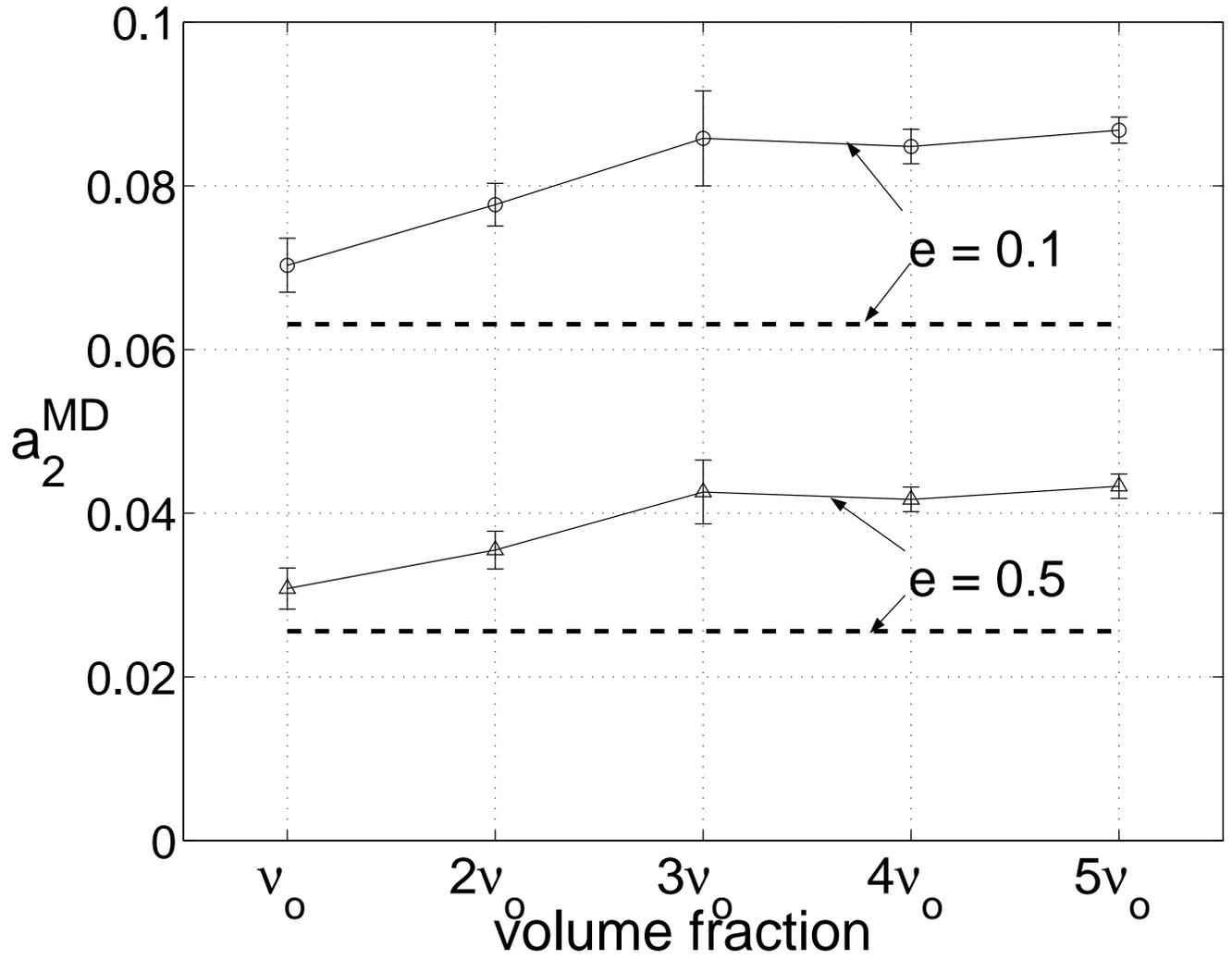}}
\smallskip
\caption{
The values of $a_2^{MD}$ obtained in the simulation as a function
of density for two coefficients of restitution, $e = 0.1$ and $e = 0.5$.
Dashed lines are the predictions of the inelastic hard sphere
kinetic theory, which do not depend on the density.
}
\end{figure}
\pagebreak

\begin{figure}
\epsfxsize=.98\columnwidth
\centerline{\epsffile{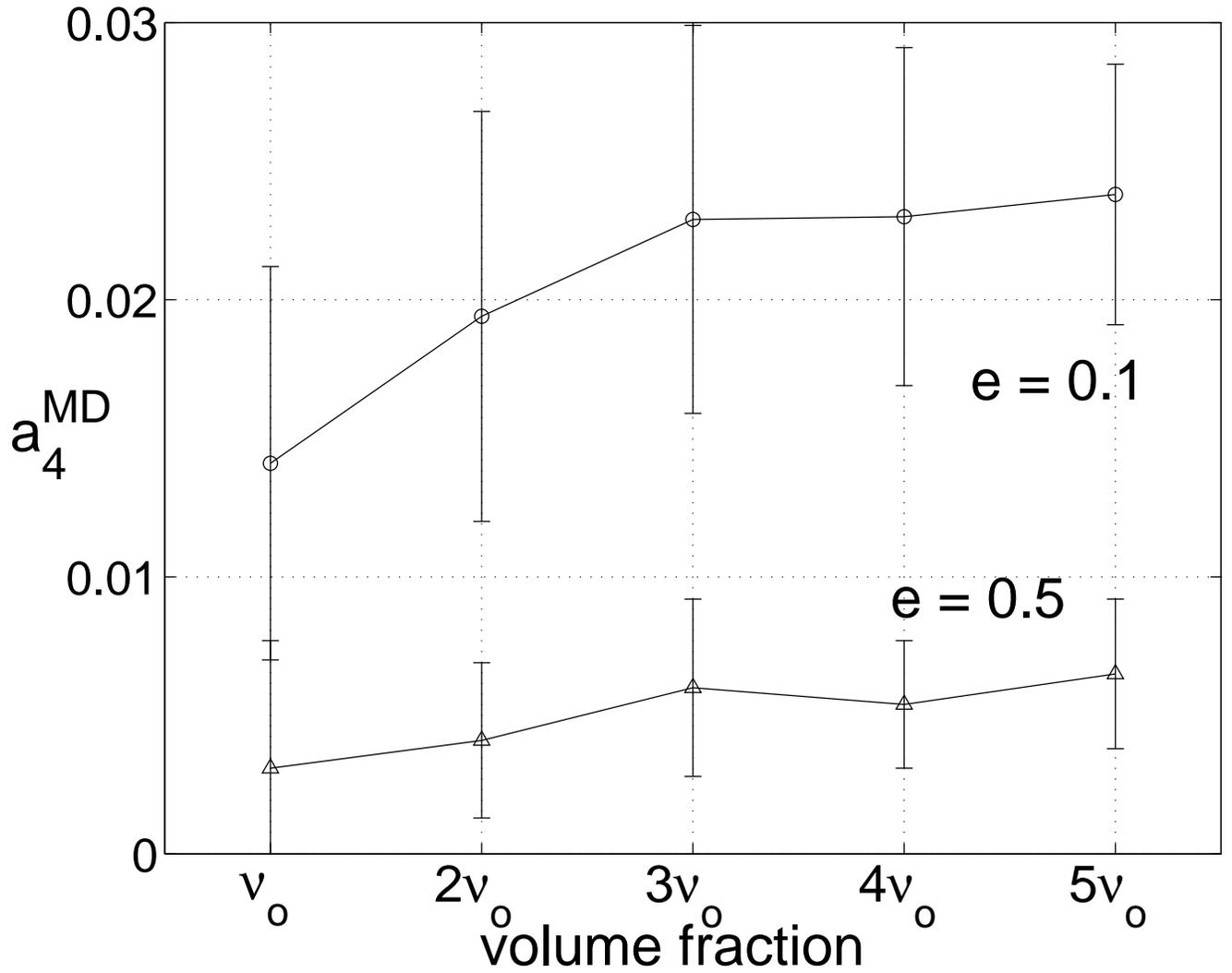}}
\smallskip
\caption{
The density dependence of $a_4^{MD}$ for two coefficients of restitution,
$e = 0.1$ and $e = 0.5$, obtained in the simulation.
Predictions are not available, because they have not been
calculated in the theoretical studies.
Error bars for these data are bigger than those for $a_2^{MD}$ (Fig. 7),
because in calculating $a_4^{MD}$, higher velocity data have more weight
exhibiting stronger fluctuations.
}
\end{figure}
\pagebreak

\begin{figure}
\epsfxsize=.98\columnwidth
\centerline{\epsffile{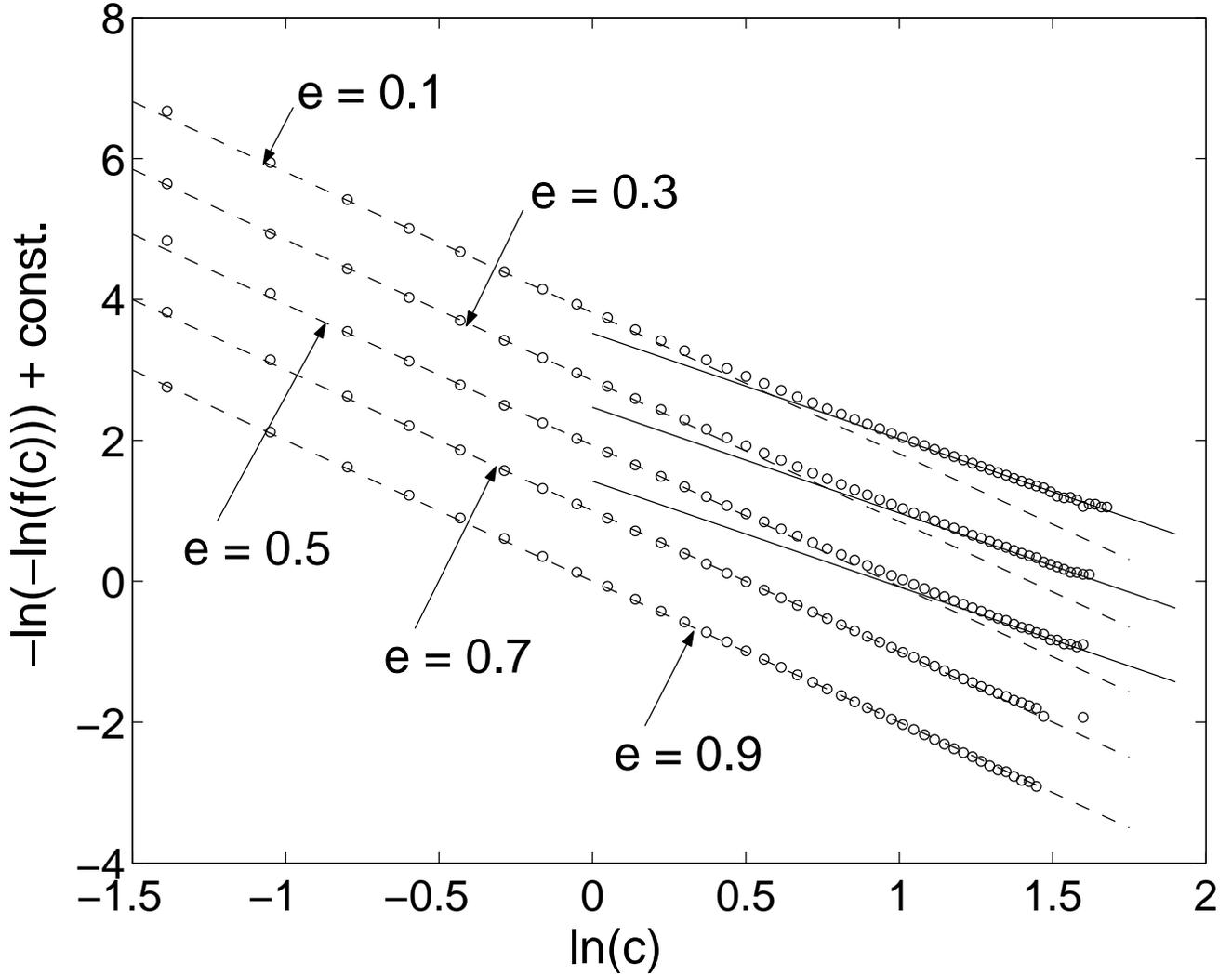}}
\smallskip
\caption{
A crossover behavior of the velocity distribution function
${\tilde f}^s_{(MD)}(c)$ (open circles) from $\sim \exp(-{\cal A}c^2)$
(compared with dashed lines) to $\sim \exp(-{\cal A}'c^{3/2})$
(compared with solid lines) is observed
for higher inelasticities ($e < 0.5$).
The volume fraction is $5\nu_o$, and
${\cal A}$ and ${\cal A}'$ are arbitrary constants.
}
\end{figure}
\pagebreak

\begin{figure}
\epsfxsize=.98\columnwidth
\centerline{\epsffile{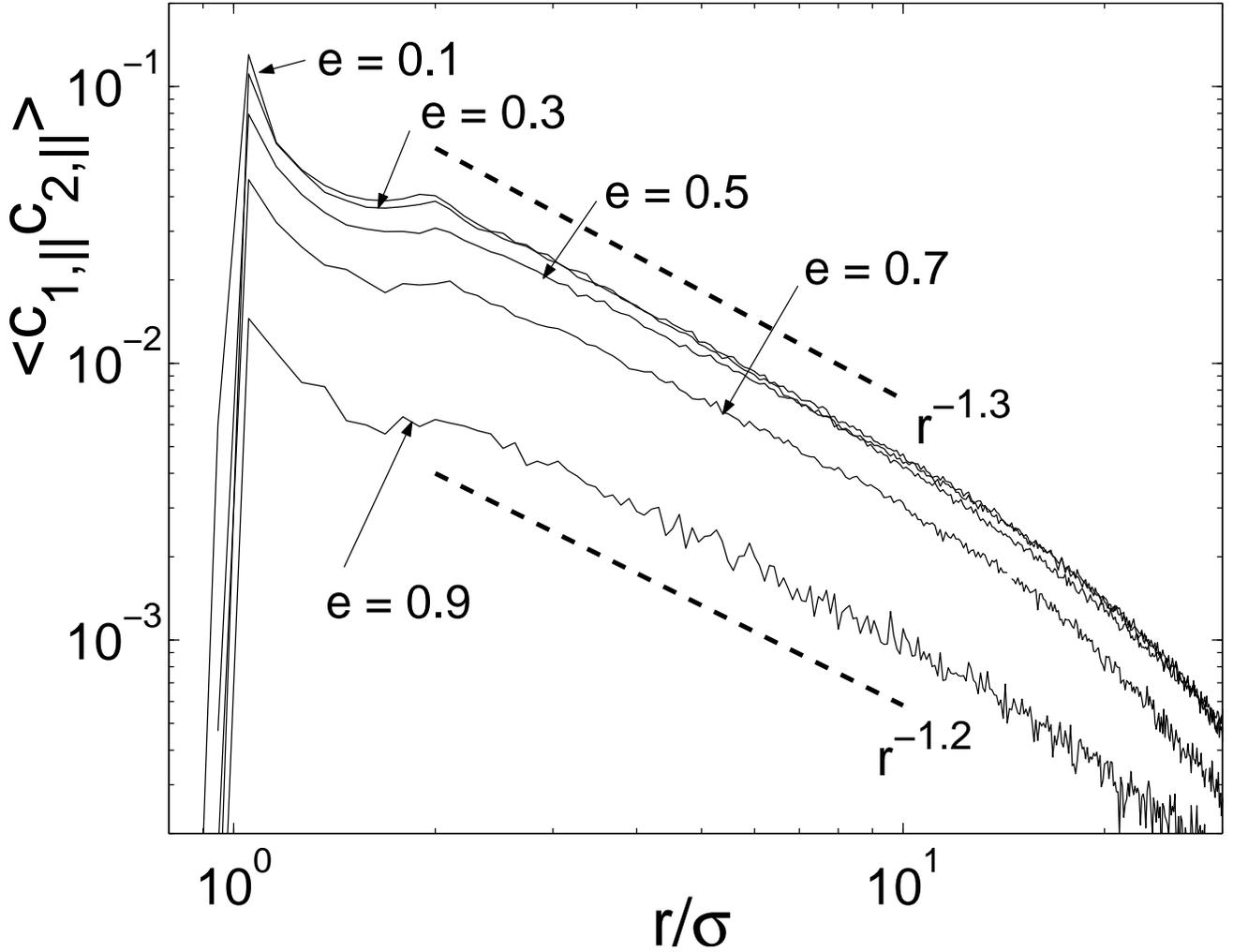}}
\smallskip
\caption{
The parallel velocity correlations for various coefficients of restitution.
The volume fraction is $5\nu_o$.
Dashed lines are the curves following the power law,
which are included for comparison.
The curves deviate from the power law for $r/\sigma > 20$,
because of the finite system size effect.
Error bars are not shown for clarity.
}
\end{figure}
\pagebreak

\begin{figure}
\epsfxsize=.98\columnwidth
\centerline{\epsffile{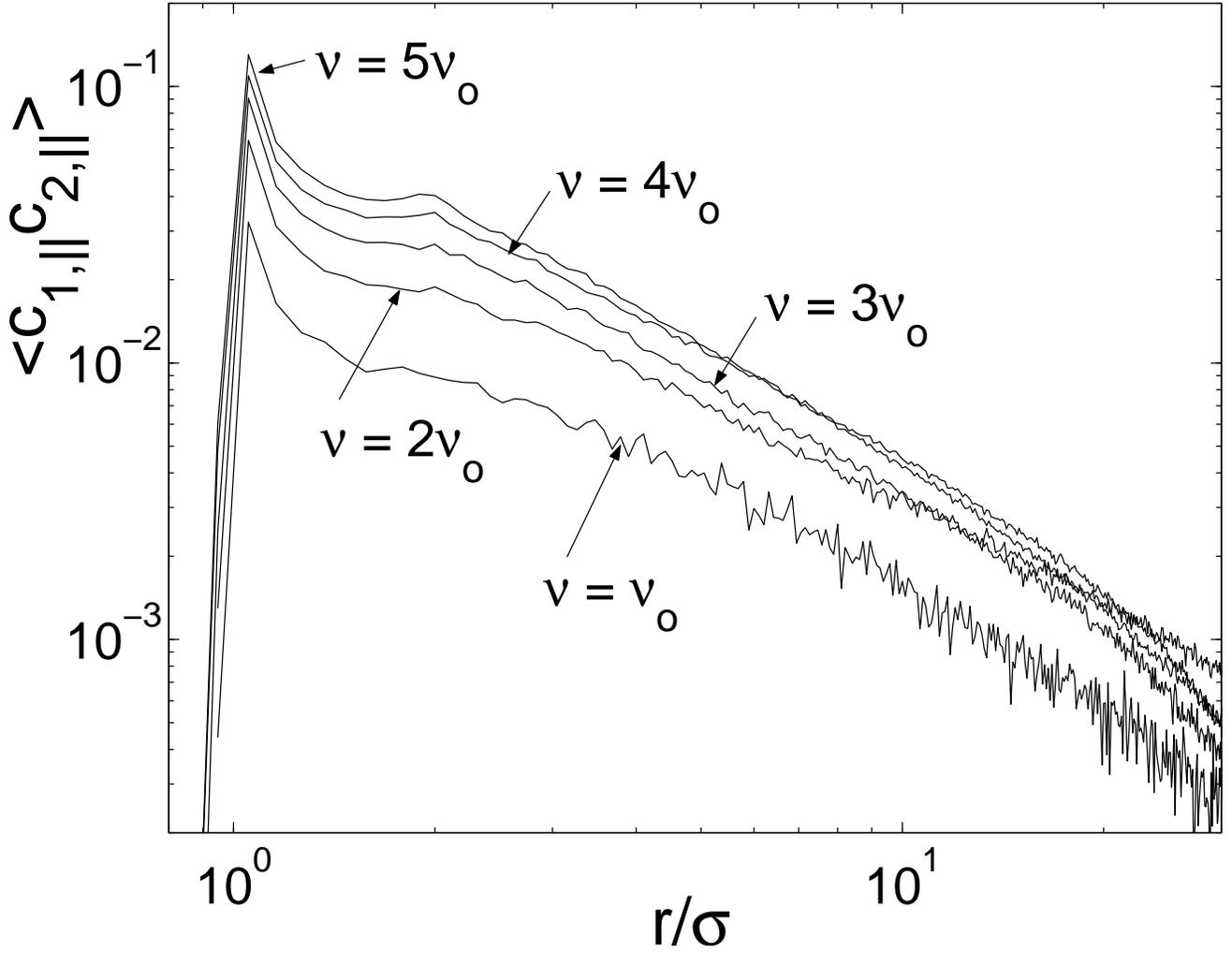}}
\smallskip
\caption{
The parallel velocity correlations for various densities.
The coefficient of restitution is $0.1$.
They have the same tendency for other coefficients of restitution
(not shown here).
}
\end{figure}
\pagebreak

\begin{figure}
\epsfxsize=.98\columnwidth
\centerline{\epsffile{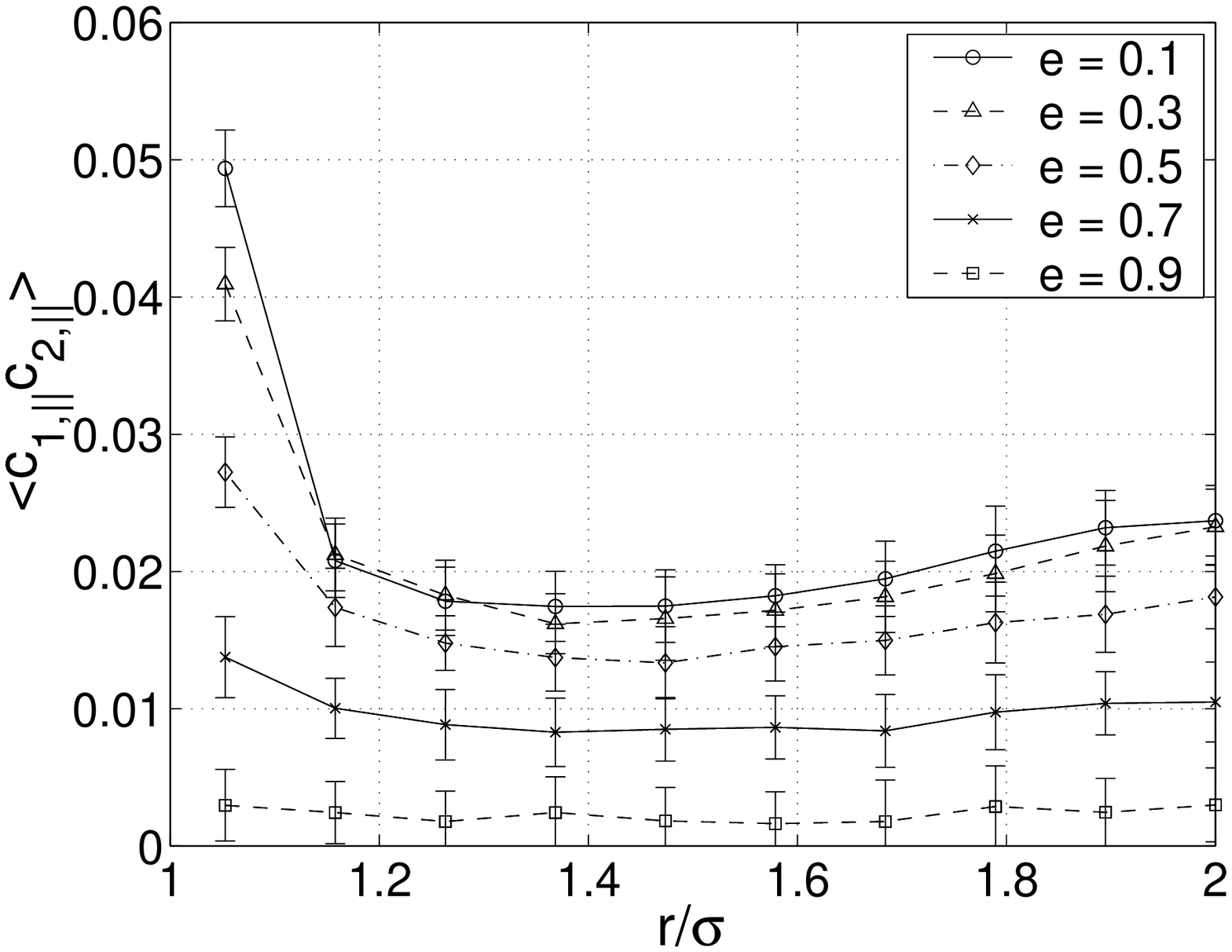}}
\smallskip
\caption{
The short-range pre-collisional parallel velocity correlations
for various coefficients of restitution.
The volume fraction is $5\nu_o$.
}
\end{figure}
\pagebreak

\begin{figure}
\epsfxsize=.98\columnwidth
\centerline{\epsffile{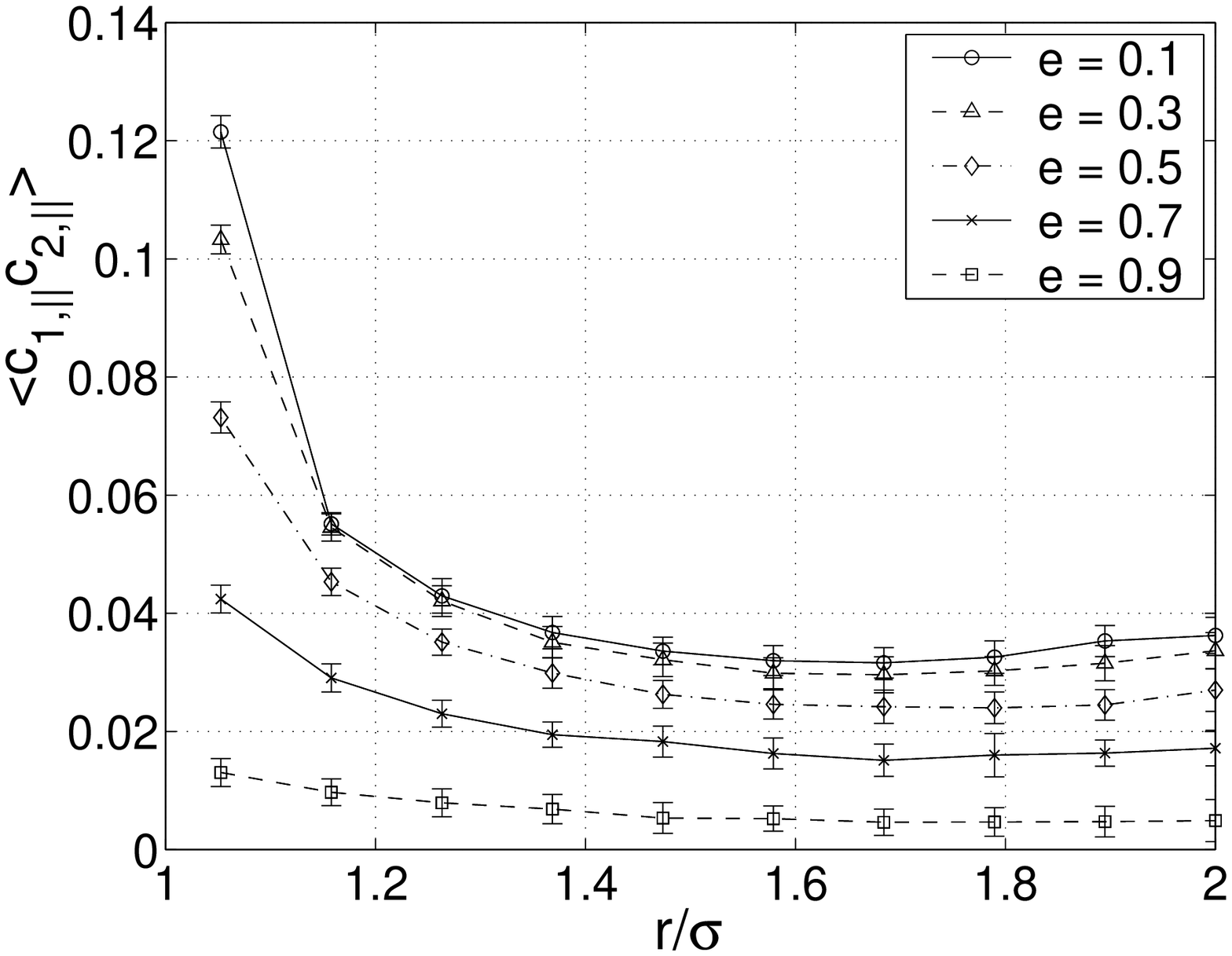}}
\smallskip
\caption{
The short-range post-collisional parallel velocity correlations
for various coefficients of restitution.
The volume fraction is $5\nu_o$.
}
\end{figure}
\pagebreak

\begin{figure}
\epsfxsize=.98\columnwidth
\centerline{\epsffile{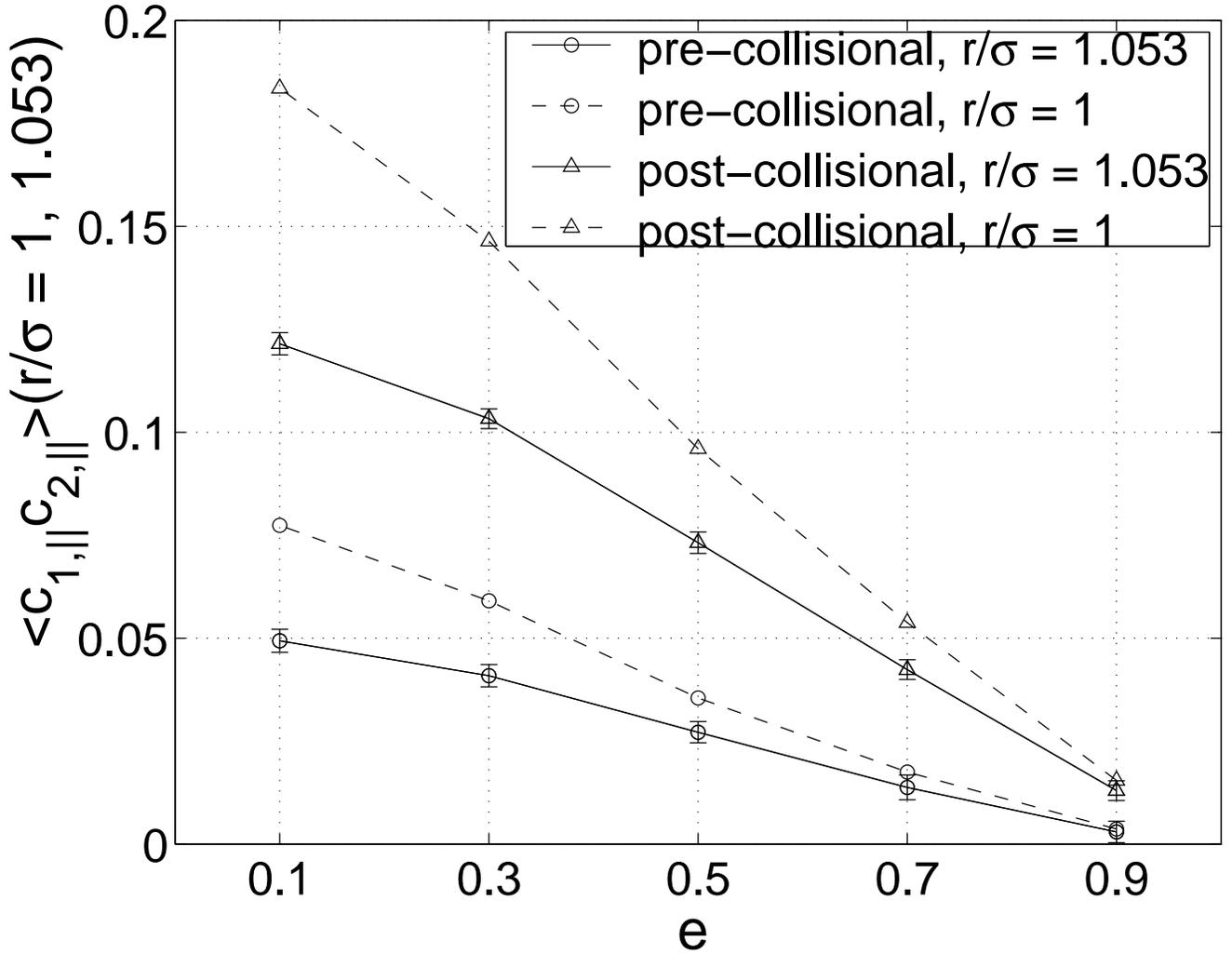}}
\smallskip
\caption{
The pre- and post-collisional parallel velocity correlations
at $r/\sigma = 1.053$ (solid lines) and estimated values at contact,
$r/\sigma = 1$ (dashed lines), as a function of the coefficient of restitution.
The volume fraction is $5\nu_o$.
The values at contact are obtained by extrapolating the data in Fig. 12
and Fig. 13, using fifth order polynomials.
These are extrapolated values from the average values,
and the error bars are not systematically determined.
The error may be of the same order as the values at $r/\sigma = 1.053$
in Fig. 12 and Fig. 13.
}
\end{figure}
\pagebreak

\begin{figure}
\epsfxsize=.98\columnwidth
\centerline{\epsffile{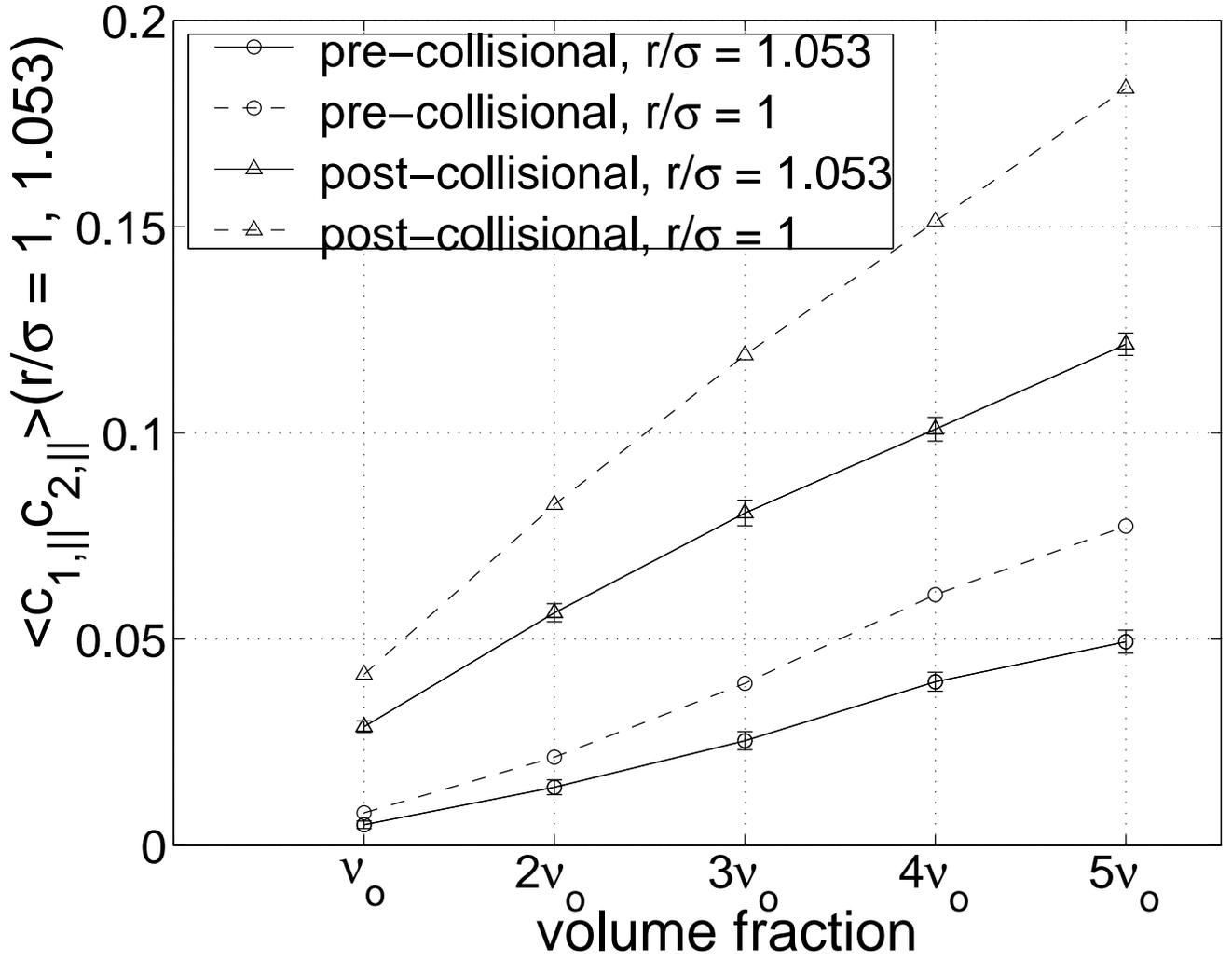}}
\smallskip
\caption{
The pre- and post-collisional parallel velocity correlations
at $r/\sigma = 1.053$ (solid lines) and estimated values at contact,
$r/\sigma = 1$ (dashed lines), as a function of the density.
The coefficient of restitution is $0.1$.
}
\end{figure}

\end{document}